\documentclass[12pt]{article}

\begin{document}

\author{C. Bizdadea\thanks{%
e-mail address: bizdadea@central.ucv.ro}, E. M. Cioroianu\thanks{%
e-mail address: manache@central.ucv.ro}, I. Negru\thanks{%
e-mail address: inegru@central.ucv.ro}, S. O. Saliu\thanks{%
e-mail address: osaliu@central.ucv.ro} \\
Faculty of Physics, University of Craiova\\
13 A. I. Cuza Str., Craiova RO-1100, Romania}
\title{Lagrangian interactions within a special class of covariant mixed-symmetry
type tensor gauge fields}
\maketitle

\begin{abstract}
Consistent non-trivial interactions within a special class of
covariant mixed-symmetry type tensor gauge fields of degree three
are constructed from the deformation of the solution to the master
equation combined with specific cohomological techniques. In
spacetime dimensions strictly greater than four, the only
consistent interaction terms are those gauge invariant under the
original symmetry. Only in four spacetime dimensions the gauge
symmetry is found to be deformed.

\end{abstract}

\section{Introduction}

The cohomological development of the Becchi-Rouet-Stora-Tyutin
symmetry, allowed, among others, the determination to be made of
the general solution to the Wess-Zumino consistency condition
\cite{1and2}--\cite{10and23}, of the general form of the
counterterms involved with the renormalization of gauge invariant
operators (Kluberg-Stern and Zuber conjecture)
\cite{6and23}--\cite {12and3}, the cohomological approach to
global symmetries and conservation laws in classical gauge
theories \cite{13and4}--\cite{16and4}, the reformulation of the
problem of constructing consistent interactions that can be
introduced in gauge-invariant theories as a problem of deformation
of the solution to the classical master equation
\cite{17and5}--\cite{21and5}, etc. In particular, the
cohomological reformulation of consistent interactions in theories
with gauge symmetries has led to important results, such as the
impossibility of cross-interactions in multigraviton theories
\cite{22and6}--\cite{24and6}, or the existence of the
Seiberg-Witten map in noncommutative field theories whose
commutative versions allow rigid (i.e. non-deformable) gauge
symmetries \cite{25and7}--\cite{27and7}. The fundamental algebraic
structure on which the BRST symmetry is based is the graded
differential complex endowed with a second-order nilpotent
differential, together with the (local) cohomology of this
differential. Lately, the usual differential tools have been
developed to cover generalized differential $N$-complexes
\cite{28and8}--\cite{29and810} for irreducible tensor gauge fields
of mixed Young symmetry type, endowed with higher-order nilpotent
operators. In this context, the generalized Poincar\'{e} lemma,
that governs the cohomology of $N$-complexes related to tensor
fields of mixed symmetry type, has been formulated and proved.
This modern differential algebraic setting helped at solving some
nice problems, like, for instance, the interpretation of the
construction of the Pauli-Fierz theory \cite{30and9}, the dual
formulation of linearized gravity \cite{31and9}--\cite{31and11},
or the impossibility of consistent cross-interactions in the dual
formulation of linearized gravity \cite {31and11}.

In this paper we investigate all consistent Lagrangian interactions that can
be added to a special class of covariant mixed-symmetry type tensor gauge
fields of degree three that transform according to a reducible
representation of the Lorentz group. Our main results are:

\noindent (i) in spacetime dimensions strictly greater than four,
the only consistent interactions are gauge invariant terms, that
do not deform the gauge symmetry;

\noindent (ii) in four dimensions, there appear non-trivial
consistent deformations that modify the gauge transformations and
their reducibility, but not the gauge algebra.

Our strategy goes as follows. Initially, we generate the
antibracket-antifield BRST symmetry of the uncoupled model with
covariant mixed-symmetry type tensor fields of degree three in an
arbitrary spacetime dimension. Next, we apply the deformation
procedure and compute the first-order deformation of the solution
to the master equation, whose integrand belongs to the
zeroth-order local cohomology of the free BRST differential
$H^{0}\left( s|d\right) $, where $s$ and $d$ mean the BRST,
respectively, the exterior spacetime differentials. Further, we
investigate the higher-order deformations. Finally, we analyze the
Lagrangian description of the interacting theory, namely, the
Lagrangian action, the deformed gauge transformations, and their
tensor structure. In $D>4$ spacetime dimensions, the first-order
deformation reduces to its antighost number zero component, while
all higher-order deformations can be taken to
vanish. Thus, the original gauge symmetry is not deformed in this case. For $%
D=4$, the first-order deformation is parameterized in terms of two
arbitrary constants. Its consistency requires that one of the
constants must vanish. Under these circumstances, the deformation
of the solution to the master equation that is consistent at all
orders in the coupling constant ends at the first order. It is
local, manifestly Lorentz covariant, and, most important,
non-trivial. It follows that the gauge transformations of the
mixed-symmetry type tensor fields are truly deformed, although
their gauge algebra remains abelian. The interacting model is also
first-order reducible, but the reducibility relations hold
on-shell, in contrast to the starting uncoupled theory.

This paper is organized into nine sections. Section 2 introduces
the Lagrangian model in an arbitrary spacetime dimension. Section
3 is devoted to the BRST symmetry of the uncoupled theory. In
Sect. 4 we briefly review the Lagrangian mechanism for
constructing consistent interactions from the point of view of the
antibracket-antifield BRST method. Section 5 focuses on the
deformation procedure in more that four spacetime dimensions and
argues that the only consistent interactions simply add
gauge-invariant terms to the starting action. In Sect. 6 we
restrict ourselves to the four-dimensional case, and reconsider
the uncoupled model in first-order form. In this context, in Sect.
7 we solve the main equations of the antibracket-antifield
deformation procedure, and show that there are consistent
deformations of the solution to the classical master equation at
order one in the coupling constant. We investigate the conditions
that must be fulfilled such that these are second- and
higher-order consistent. In Sect. 8 we identify the Lagrangian
version of the interacting model in four dimensions, and analyze
the tensor structure of the new gauge theory. Section 9 ends the
paper with some conclusions.

\section{Introducing the free model}

We start from the Lagrangian action for a special class of
covariant mixed-symmetry type tensor gauge fields of degree three:
\begin{equation}
S_{0}\left[ A_{\alpha \beta }^{\;\;\;(\sigma )}\right] =-\frac{1}{12}\int
d^{D}x\,F_{\alpha \beta \gamma }^{\;\;\;\;(\sigma )}F_{\;\;\;\;(\sigma
)}^{\alpha \beta \gamma },  \label{fcol1}
\end{equation}
where the fields are only antisymmetric in their first two indices $%
A_{\alpha \beta }^{\;\;\;(\sigma )}=-A_{\beta \alpha
}^{\;\;\;(\sigma )}$. The tensor fields $A_{\alpha \beta
}^{\;\;\;(\sigma )}$ can be regarded as being described by a Young
diagram with three cells and two columns. However, they \textit{do
not satisfy} the identity $A_{\left[ \alpha \beta (\sigma )\right]
}=0$, associated with the Young symmetrizer of the corresponding
diagram. Actually, they transform according to a reducible
representation of the Lorentz group. This is the reason why we use
the notation $A_{\alpha \beta (\sigma )}$ instead of the standard
one, $A_{\alpha \beta |\sigma }$. Spacetime indices are raised and
lowered with the flat
Minkowskian metric of ``mostly plus'' signature in $D$ dimensions ($D\geq 4$%
): $-+++\cdots $. We define the field strength components in the
usual manner:
\begin{equation}
F_{\alpha \beta \gamma }^{\;\;\;\;(\sigma )}=\partial _{\alpha }A_{\beta
\gamma }^{\;\;\;(\sigma )}+\partial _{\beta }A_{\gamma \alpha
}^{\;\;\;(\sigma )}+\partial _{\gamma }A_{\alpha \beta }^{\;\;\;(\sigma
)}\equiv \partial _{\left[ \alpha \right. }A_{\left. \beta \gamma \right]
}^{\;\;\;\;(\sigma )},  \label{fcol2}
\end{equation}
where we choose the convention that $\left[ \alpha \beta \cdots
\right] $ signifies plain antisymmetry with respect to the indices
between brackets, without additional numerical factors. The tensor
fields $A_{\alpha \beta }^{\;\;\;(\sigma )}$ can also be viewed in
some sense as a collection of two-forms, where the collection
index is spacetime-like. This motivates the form (\ref{fcol2}) of
their field strengths, as well as the expression (\ref {fcol1}) of
the Lagrangian action. It is widely known that gauge theories
involving antisymmetric tensor fields are important due to their
connection with string theory and supergravity models
\cite{string1}--\cite{string6}. In particular, interacting
two-forms (described by the Freedman-Townsend model) play a
special role due on the one hand to its link with Witten's string
theory \cite{string7} and, on the other hand, to its equivalence
to the nonlinear sigma model \cite{string8}. The interacting
theory to be derived by us in Sect. 8 resembles in a way the
Freedman-Townsend
model, but clearly exhibits new features. The gauge invariances of action (%
\ref{fcol1}) are given by
\begin{equation}
\delta _{\epsilon }A_{\alpha \beta }^{\;\;\;(\sigma )}=\partial _{\left[
\alpha \right. }\epsilon _{\left. \beta \right] }^{\;\;\;(\sigma )}\equiv
\stackrel{(0)}{R}_{\alpha \beta \;\;\;\;(\rho )}^{\;\;\;(\sigma )\gamma
}\epsilon _{\gamma }^{\;(\rho )},  \label{fcol3}
\end{equation}
with $\epsilon _{\gamma }^{\;(\rho )}$ bosonic gauge parameters,
that are tensors of degree two with no symmetry, such that the
gauge generators in condensed De Witt notation are expressed
through
\begin{equation}
\stackrel{(0)}{R}_{\alpha \beta \;\;\;\;(\rho )}^{\;\;\;(\sigma )\gamma
}=\delta _{\rho }^{\sigma }\partial _{\left[ \alpha \right. }\delta _{\left.
\beta \right] }^{\gamma }.  \label{fcol4}
\end{equation}
The above gauge transformations are abelian and turn out to be
off-shell first-stage reducible, the accompanying reducibility
relations and functions respectively reading
\begin{equation}
\stackrel{(0)}{R}_{\alpha \beta \;\;\;\;(\rho )}^{\;\;\;(\sigma )\gamma }%
\stackrel{(0)}{Z}_{\gamma \;\;\;(\tau )}^{\;(\rho )}=0,  \label{fcol5}
\end{equation}
\begin{equation}
\stackrel{(0)}{Z}_{\gamma \;\;\;(\tau )}^{\;(\rho )}=\delta _{\tau }^{\rho
}\partial _{\gamma }.  \label{fcol6}
\end{equation}

In consequence, we deal with a usual free linear gauge theory of
Cauchy order three (being local, polynomial in the fields and
their derivatives, and satisfying the standard regularity
conditions \cite{32and12}, whose gauge algebra is not open, and
with a finite reducibility order). As the number of physical
degrees of freedom carried by a tensor field $A_{\alpha \beta
}^{\;\;\;(\sigma )}$ is equal to $D\left( D-2\right) \left(
D-3\right) /2$, the subsequent analysis is meaningful only for
$D\geq 4$.

\section{BRST symmetry of the free theory}

We are interested in constructing the consistent Lagrangian
(self)interactions that can be added to the action (\ref{fcol1})
without changing the content of the field spectrum or the number
of independent gauge symmetries. In view of this, we apply the
general rules of the antibracket-antifield deformation procedure
based on (co)homological techniques \cite{17and5}, \cite{21and5}.

The first step in the development of the antibracket-antifield deformation
approach consists in generating the Lagrangian BRST symmetry of the free
model under study. The Lagrangian BRST complex is organized into the
field/ghost, respectively, antifield spectrum
\begin{equation}
\Phi ^{\Delta }=\left( A_{\alpha \beta }^{\;\;\;(\sigma )},\eta _{\alpha
}^{\;(\sigma )},C^{(\sigma )}\right) ;\;\Phi _{\Delta }^{*}=\left(
A_{\;\;\;\;(\sigma )}^{*\alpha \beta },\eta _{\;\;\;(\sigma )}^{*\alpha
},C_{(\sigma )}^{*}\right) ,  \label{fcol12}
\end{equation}
with the Grassmann parities
\begin{equation}
\varepsilon \left( A_{\alpha \beta }^{\;\;\;(\sigma )}\right) =0=\varepsilon
\left( C^{(\sigma )}\right) ,\;\varepsilon \left( \eta _{\alpha }^{\;(\sigma
)}\right) =1,  \label{fcol13}
\end{equation}
\begin{equation}
\varepsilon \left( \Phi _{\Delta }^{*}\right) =\left( \varepsilon \left(
\Phi ^{\Delta }\right) +1\right) \;\mathrm{mod}\;2.  \label{fcol15}
\end{equation}
While the ghosts $\eta _{\alpha }^{\;(\sigma )}$ are due to the
gauge symmetry, the ghosts for ghosts $C^{(\sigma )}$ are required
by the first-order reducibility relations. The Lagrangian BRST
symmetry acts like a differential $s$ ($s^{2}=0$), which we assume
to behave like a right derivation. Since the gauge algebra is
abelian and the reducibility functions are field independent, it
follows that $s$ reduces to the sum of the Koszul-Tate
differential $\delta $ and the exterior longitudinal derivative
$\gamma $ only,
\begin{equation}
s=\delta +\gamma ,  \label{fcol16}
\end{equation}
that are respectively graded in terms of the antighost number ($\mathrm{agh}$%
) and the pure ghost number ($\mathrm{pgh}$).\footnote{%
For more general gauge theories, $s$ has a richer structure than in (\ref
{fcol16}), the addition of supplementary operators $%
\stackrel{(k)}{s}$, with $\mathrm{agh}\left(
\stackrel{(k)}{s}\right) =k>0$ being necessary, in order to ensure
the second-order nilpotency of $s$.} While the Koszul-Tate
differential ($\delta ^{2}=0$, $\mathrm{agh}\left( \delta \right)
=-1$, $\mathrm{pgh}\left( \delta \right) =0$) realizes a
resolution of smooth functions defined on the stationary surface
of the field equations,
the exterior longitudinal derivative ($\mathrm{pgh}\left( \gamma \right) =1$%
, $\mathrm{agh}\left( \gamma \right) =0$) anticommutes with
$\delta $ and turns out to be a true differential in the
particular case of the model under study $\left( \gamma
^{2}=0\right) $. Its cohomological space at pure ghost number zero
computed in the homology of $\delta $, $H^{0}\left( \gamma
|H_{*}\left( \delta \right) \right) $, is given by the algebra of
the Lagrangian physical observables, and is in the meantime
isomorphic to the zeroth order cohomological space of $s$,
$H^{0}\left( s\right) $, that contains the
so-called BRST observables \cite{33and13}--\cite{35and13}. The degrees ($%
\mathrm{agh}$) and ($\mathrm{pgh}$) of the BRST generators (\ref{fcol12})
are given by
\begin{equation}
\mathrm{agh}\left( \Phi ^{\Delta }\right) =0,\;\mathrm{agh}\left(
A_{\;\;\;\;(\sigma )}^{*\alpha \beta }\right) =1,\;\mathrm{agh}\left( \eta
_{\;\;\;(\sigma )}^{*\alpha }\right) =2,\;\mathrm{agh}\left( C_{(\sigma
)}^{*}\right) =3,  \label{fcol17}
\end{equation}
\begin{equation}
\mathrm{pgh}\left( \Phi _{\Delta }^{*}\right) =0,\;\mathrm{pgh}\left(
A_{\alpha \beta }^{\;\;\;(\sigma )}\right) =0,\;\mathrm{pgh}\left( \eta
_{\alpha }^{\;(\sigma )}\right) =1,\;\mathrm{pgh}\left( C^{(\sigma )}\right)
=2,  \label{fcol18}
\end{equation}
while the actions of $\delta $ and $\gamma $ read
\begin{equation}
\delta \Phi ^{\Delta }=0,\;\delta A_{\;\;\;\;(\sigma )}^{*\alpha \beta }=-%
\frac{\delta S_{0}^{L}}{\delta A_{\alpha \beta }^{\;\;\;(\sigma )}}\equiv -%
\frac{1}{2}\partial _{\gamma }F_{\;\;\;\;(\sigma )}^{\gamma \alpha \beta },
\label{fcol19}
\end{equation}
\begin{equation}
\delta \eta _{\;\;\;(\sigma )}^{*\alpha }=-2\partial _{\beta
}A_{\;\;\;\;(\sigma )}^{*\beta \alpha },\;\delta C_{(\sigma )}^{*}=\partial
_{\alpha }\eta _{\;\;\;(\sigma )}^{*\alpha },  \label{fcol20}
\end{equation}
\begin{equation}
\gamma \Phi _{\Delta }^{*}=0,\;\gamma A_{\alpha \beta }^{\;\;\;(\sigma
)}=\partial _{\left[ \alpha \right. }\eta _{\left. \beta \right]
}^{\;\;\;(\sigma )},\;\gamma \eta _{\alpha }^{\;(\sigma )}=\partial _{\alpha
}C^{(\sigma )},\;\gamma C^{(\sigma )}=0.  \label{fcol21}
\end{equation}
The overall degree of the BRST complex is named ghost number ($\mathrm{gh}$)
and is defined like $\mathrm{gh}=\mathrm{pgh}-\mathrm{agh}$, such that $%
\mathrm{gh}\left( s\right) =1$. An important property of the Lagrangian BRST
symmetry is that it is canonically generated in a structure named
antibracket and usually denoted by $\left( ,\right) $ \cite{rusi1}--\cite
{35and13} via a generator $\stackrel{(0)}{S}$:%
\begin{equation}
s\cdot =\left( \cdot ,\stackrel{(0)}{S}\right) .  \label{fcol21a}
\end{equation}
The fields/ghosts are decreed conjugated in the antibracket with
the associated antifields:
\begin{equation}
\left( \Phi ^{\Delta },\Phi _{\Delta ^{\prime }}^{*}\right) =\delta _{\Delta
^{\prime }}^{\Delta }.  \label{fcol22}
\end{equation}
The second-order nilpotency of $s$ is equivalent to the fact that $\stackrel{%
(0)}{S}$ is solution to the classical master equation:
\begin{equation}
\left( \stackrel{(0)}{S},\stackrel{(0)}{S}\right) =0,\;\varepsilon \left(
\stackrel{(0)}{S}\right) =0,\;\mathrm{gh}\left( \stackrel{(0)}{S}\right) =0,
\label{fcol23}
\end{equation}
with some boundary conditions. In the case of the model investigated here,
with the help of the relations (\ref{fcol16}) and (\ref{fcol19}--\ref
{fcol21a}), one finds that the complete solution to the master equation can
be written in the form
\begin{equation}
\stackrel{(0)}{S}=S_{0}\left[ A_{\alpha \beta }^{\;\;\;(\sigma )}\right]
+\int d^{D}x\left( A_{\;\;\;\;(\sigma )}^{*\alpha \beta }\partial _{\left[
\alpha \right. }\eta _{\left. \beta \right] }^{\;\;\;(\sigma )}+\eta
_{\;\;\;(\sigma )}^{*\alpha }\partial _{\alpha }C^{(\sigma )}\right) ,
\label{fcol24}
\end{equation}
and we observe that it contains pieces of antighost number equal
to zero, one, and two, respectively.

\section{Brief review of antibracket-antifield deformation procedure}

In order to develop the general approach to the problem of consistent
interactions in gauge theories from the perspective of the
antibracket-antifield deformation procedure, we briefly recall some basic
results \cite{17and5}, \cite{21and5}. Assume that
\begin{equation}
\bar{S}_{0}\left[ A_{\alpha \beta }^{\;\;\;(\sigma )}\right]
=S_{0}\left[ A_{\alpha \beta }^{\;\;\;(\sigma )}\right] +g\int
d^{D}x\,a_{0}+O\left( g^{2}\right)  \label{fcol26}
\end{equation}
defines a consistent deformation of the action (\ref{fcol1}), with
deformed gauge symmetry
\begin{equation}
\bar{\delta}_{\epsilon }A_{\alpha \beta }^{\;\;\;(\sigma )}=\bar{R}_{\alpha
\beta \;\;\;\;(\rho )}^{\;\;\;(\sigma )\gamma }\epsilon _{\gamma }^{\;(\rho
)},  \label{defgsym}
\end{equation}
where
\begin{equation}
\bar{R}_{\alpha \beta \;\;\;\;(\rho )}^{\;\;\;(\sigma )\gamma }=\stackrel{(0)%
}{R}_{\alpha \beta \;\;\;\;(\rho )}^{\;\;\;(\sigma )\gamma }+g\stackrel{(1)}{%
R}_{\alpha \beta \;\;\;\;(\rho )}^{\;\;\;(\sigma )\gamma }+O\left(
g^{2}\right) .  \label{fcol27}
\end{equation}
Consistency means that $\bar{S}_{0}$ is fully invariant under the
gauge
symmetry $\bar{\delta}_{\epsilon }$ (at any order in the coupling constant $%
g $), $\bar{\delta}_{\epsilon }\bar{S}_{0}=0$. Moreover, we add
the demand that the number of independent deformed gauge
symmetries should remain unchanged with respect to the free
theory, which implies the existence of some first-order
reducibility functions for the interacting theory,
\begin{equation}
\bar{Z}_{\gamma \;\;\;(\tau )}^{\;(\rho )}=\stackrel{(0)}{Z}_{\gamma
\;\;\;(\tau )}^{\;(\rho )}+g\stackrel{(1)}{Z}_{\gamma \;\;\;(\tau
)}^{\;(\rho )}+O\left( g^{2}\right) ,  \label{fcol28}
\end{equation}
such that the new reducibility relations may now hold on-shell, i.e., on the
stationary surface of field equations for $\bar{S}_{0}$:%
\begin{equation}
\bar{R}_{\alpha \beta \;\;\;\;(\rho )}^{\;\;\;(\sigma )\gamma }\bar{Z}%
_{\gamma \;\;\;(\tau )}^{\;(\rho )}\approx 0.  \label{fcol29}
\end{equation}
It is possible to reformulate more economically this problem in terms of the
solution to the master equation. The key observation on which this approach
relies is that a consistent deformation of the free action (\ref{fcol1}) and
of its gauge symmetries (\ref{fcol3}) defines a deformation of the solution (%
\ref{fcol24}) of the master equation that preserves both the
master equation and the field/ghost-antifield spectrum. Indeed, if
the interactions can be consistently constructed, then the
solution (\ref{fcol24}) of the master equation for the free theory
can be deformed into the solution $\bar{S}$ of the master equation
for the interacting theory, thus:
\begin{equation}
\stackrel{(0)}{S}\rightarrow \bar{S}=\stackrel{(0)}{S}+g\stackrel{(1)}{S}%
+g^{2}\stackrel{(2)}{S}+\cdots ,  \label{fcol25}
\end{equation}
\begin{equation}
\left( \stackrel{(0)}{S},\stackrel{(0)}{S}\right) =0\rightarrow \left( \bar{S%
},\bar{S}\right) =0.  \label{fcol7}
\end{equation}
The master equation $\left( \bar{S},\bar{S}\right) =0$ guarantees that the
consistency requirements on $\bar{S}_{0}$, $\bar{R}_{\alpha \beta
\;\;\;\;(\rho )}^{\;\;\;(\sigma )\gamma }$ and $\bar{Z}_{\gamma \;\;\;(\tau
)}^{\;(\rho )}$ are fulfilled. The main advantage in reformulating the
problem of consistent interactions as the problem of deforming the master
equation is that we can make use of the cohomological techniques of the
deformation theory. The master equation for $\bar{S}$ splits according to
the deformation parameter $g$ as
\begin{equation}
g^{0}:\left( \stackrel{(0)}{S},\stackrel{(0)}{S}\right) =0,  \label{fcol9a}
\end{equation}
\begin{equation}
g^{1}:2\left( \stackrel{(1)}{S},\stackrel{(0)}{S}\right) =0,  \label{fcol10}
\end{equation}
\begin{equation}
g^{2}:2\left( \stackrel{(2)}{S},\stackrel{(0)}{S}\right) +\left( \stackrel{%
(1)}{S},\stackrel{(1)}{S}\right) =0,  \label{fcol11}
\end{equation}
\[
\vdots
\]
Equation (\ref{fcol9a}) is checked by assumption, while (%
\ref{fcol10}) stipulates that the first-order deformation is a
cocycle of the free BRST differential (\ref{fcol21a}),
\begin{equation}
s\stackrel{(1)}{S}=0.  \label{fcol30}
\end{equation}
However, only cohomologically non-trivial solutions to
(\ref{fcol10}) should be taken into account, since trivial
(BRST-exact) ones can be eliminated by a (in general non-linear)
field redefinition \cite{21and5}. In this way, we conclude that
the non-trivial deformations $\stackrel{(1)}{S}$ are determined by
the zeroth-order cohomological space $H^{0}\left( s\right) $ of
the undeformed theory, which is generically non-empty due to its
isomorphism to
the space of physical observables of the free theory. The next equation, (%
\ref{fcol11}), implies that $\left( \stackrel{(1)}{S},\stackrel{(1)}{S}%
\right) $ must be trivial (BRST-exact) in the cohomology of $s$ at
ghost number one, $H^{1}\left( s\right) $. There are no
obstructions in finding solutions to the remaining equations
(\ref{fcol11}), etc. (for instance, see \cite{21and5})) as long as
no restrictions on the interactions, such as spacetime locality or
manifest Lorentz covariance, are imposed. In general, the
resulting interactions may be nonlocal, and there might even
appear obstructions if one insists on their locality. However, as
it will be seen below, in the case of the model under
consideration we obtain only local and manifestly covariant
interactions.

\section{Deformations in $D>4$}

There are three different types of deformations, namely:

\noindent (i) that modify only the Lagrangian action, but not the
gauge symmetry;

\noindent (ii) that change the gauge symmetries, but not their
algebra; and

\noindent (iii) that deform also the gauge algebra. In spacetime
dimensions strictly greater than four ($D>4$), the only consistent
non-trivial deformations for the model under study are
shown to be only of type (i), and hence they merely add to the free action (%
\ref{fcol1}) interaction terms that are invariant under the original gauge
transformations (\ref{fcol3}). This can be seen by developing the
first-order deformation of the solution to the master equation according to
the antighost number
\begin{equation}
\stackrel{(1)}{S}=\sum_{k=0}^{n}\int d^{D}x\,a_{k},\;\mathrm{gh}\left(
a_{k}\right) =0,\;\varepsilon \left( a_{k}\right) =0,\;\mathrm{agh}\left(
a_{k}\right) =k.  \label{fcoli1}
\end{equation}
The equation for $\stackrel{(1)}{S}$, namely (\ref{fcol30}) takes
the local form
\begin{equation}
s\left( \sum_{k=0}^{n}a_{k}\right) =\partial _{\mu }h^{\mu }.  \label{fcoli2}
\end{equation}
The number of terms from the expansion of $\stackrel{(1)}{S}$ is finite and
it can be shown that one can take the piece of highest antighost number to
be $\gamma $-closed, $\gamma a_{n}=0$, and so $a_{n}\in H\left( \gamma
\right) $. This can be done by using the arguments from \cite{7and23}, \cite
{23and6}, \cite{32and12}, \cite{36and14}. From the definitions (\ref{fcol19}%
--\ref{fcol21}), it is simple to see that $H\left( \gamma \right) $ is
generated by the field strength components $F_{\alpha \beta \gamma
}^{\;\;\;\;(\sigma )}$ and their derivatives, by the antifields $\Phi
_{\Delta }^{*}$ together with their derivatives, as well as by the
undifferentiated ghosts for ghosts $C^{(\sigma )}$. Since $\mathrm{gh}\left(
C^{(\sigma )}\right) =2$, it follows that $n=2m$ and
\begin{equation}
a_{n}\equiv a_{2m}=\mu ^{\rho _{1}\cdots \rho _{m}}\left( \left[ \Phi
_{\Delta }^{*}\right] ,\left[ F_{\alpha \beta \gamma }^{\;\;\;\;(\sigma
)}\right] \right) C_{(\rho _{1})}\cdots C_{(\rho _{m})},  \label{fcoli3}
\end{equation}
where the notation $f\left( \left[ q\right] \right) $ signifies
that $f$ depends on $q$ and its derivatives up to a finite order.
The spacetime derivatives of the ghosts for ghosts are exact in
$H\left( \gamma \right) $ according to the third relation in
(\ref{fcol21}), and hence trivial, such that they are discarded
from $a_{2m}$. By projecting (\ref
{fcoli2}) on antighost number $\left( 2m-1\right) $, we infer the equation $%
\delta a_{2m}+\gamma a_{2m-1}=\partial _{\mu }h^{\prime \mu }$.
Replacing the expression (\ref{fcoli3}) in the last equation, we
find the result that a
necessary condition for the existence of $a_{2m-1}$ is that the functions $%
\mu ^{\rho _{1}\cdots \rho _{m}}$ belong to the local homology of the
Koszul-Tate differential at antighost number $2m$, $\mu ^{\rho _{1}\cdots
\rho _{m}}\in H_{2m}\left( \delta |d\right) $,%
\begin{equation}
\delta \mu ^{\rho _{1}\cdots \rho _{m}}=\partial _{\mu }h^{\prime \mu \rho
_{1}\cdots \rho _{m}},\;\mathrm{agh}\left( \mu ^{\rho _{1}\cdots \rho
_{m}}\right) =2m,\;\mathrm{pgh}\left( \mu ^{\rho _{1}\cdots \rho
_{m}}\right) =0.  \label{fcoli4}
\end{equation}
But as the model under study is linear and of Cauchy order equal
to three, we have the local homology of $\delta $ vanishing for
antighost numbers greater that three, $H_{k}\left( \delta
|d\right) =0$, $k>3$. On the other hand, the last representative,
$a_{n}$, is constructed from some functions that pertain to an
even-order space $H_{2m}\left( \delta |d\right) $, such that the
first admitted value $2m\leq 3$ is $m=1$ ($n=2$). Then we can
assume that $\stackrel{(1)}{S}=\sum_{k=0}^{2}\int d^{D}x\,a_{k}$, where $%
a_{2}=\mu ^{\rho }C_{(\rho )}$, with $\mu ^{\rho }\in H_{2}\left( \delta
|d\right) $.

At this stage, we remark that the general representative of $H_{2}\left(
\delta |d\right) $ is of the form $\alpha =\lambda _{\alpha }^{\;(\sigma
)}\eta _{\;\;\;(\sigma )}^{*\alpha }$, for some constants $\lambda _{\alpha
}^{\;(\sigma )}$. Consequently, it follows that $\mu ^{\rho }=\lambda
^{\sigma }\eta _{\;\;\;(\sigma )}^{*\rho }+\tilde{\lambda}_{\alpha }\eta
^{*\alpha (\rho )}$, where $\lambda ^{\sigma }$ and $\tilde{\lambda}_{\alpha
}$ are also constant. On the other hand, the only covariant choice of these
constants is $\lambda ^{\sigma }=k_{1}\partial ^{\sigma }$ and $\tilde{%
\lambda}_{\alpha }=k_{2}\partial _{\alpha }$, which further yields
\begin{equation}
\mu ^{\rho }=k_{1}\partial ^{\sigma }\eta _{\;\;\;(\sigma )}^{*\rho
}+k_{2}\partial _{\alpha }\eta ^{*\alpha (\rho )},  \label{lala1}
\end{equation}
with $k_{1,2}$ numerical constants. The second term on the right
hand-side of (\ref{lala1}) is $\delta $-exact (see the latter
relations in (\ref {fcol20})). As one can always add a $\delta
$-exact term to the solution of
the equation $\delta \mu ^{\rho }=\partial _{\alpha }h^{\prime \alpha \rho }$%
, we have the result that the second term from the right hand-side
of (\ref{lala1}) can be removed. Then we find that
\begin{equation}
a_{2}=\partial ^{\sigma }\left( k_{1}\eta _{\;\;\;(\sigma )}^{*\rho
}C_{(\rho )}\right) -\gamma \left( k_{1}\eta _{\;\;\;(\sigma )}^{*\rho }\eta
_{\;(\rho )}^{\sigma }\right) .  \label{lala2}
\end{equation}
Since we are free to add a $\gamma $-exact term and a $\gamma
$-invariant divergence to the last representative, we get the
result that $a_{2}$ can be chosen to vanish, $a_{2}=0$. Further,
$a_{1}$ is not eligible as the last representative in
$\stackrel{(1)}{S}$ because it does not display an even pure ghost
number. As a consequence, $\stackrel{(1)}{S}$ simply reduces to
the component that is ghost and antifield independent,
$\stackrel{(1)}{S}=\int d^{D}x\,a_{0}$, where $a_{0}$ is a $\gamma
$-cocycle modulo $d$, $\gamma a_{0}=\partial _{\mu }h^{\prime
\prime \mu }$. The non-trivial solutions to these equations are of
two kinds. The first one corresponds to $h^{\prime \prime \mu }=0$
and is given by local functions that are invariant under the
original gauge transformations, which are polynomials in the field
strength components $F_{\alpha \beta \gamma }^{\;\;\;\;(\sigma )}$
and their spacetime derivatives, while the second kind is
associated with $h^{\prime \prime \mu }\neq 0$ and is spanned by
generalized Chern-Simons terms.

In conclusion, the only non-trivial first-order deformation of the
solution to the master equation for a special class of
mixed-symmetry type tensor gauge fields of degree three in
spacetime dimensions strictly greater than four is, according to
the classification from the beginning of this section, of type
(i). Beside the Lagrangian action, neither the gauge algebra, nor
the gauge transformations, nor the reducibility relations are
modified. The deformed solution $\bar{S}=\stackrel{(0)}{S}+g\int
d^{D}x\,a_{0}$ is already
consistent to all orders in the deformation parameter, so we can take $%
\stackrel{(k)}{S}=0$, $k>1$.

\section{Four-dimensional model in first-order form}

Since there are no deformations of the free solution
(\ref{fcol24}) to the master equation in more that four spacetime
dimensions that truly deform the gauge transformations and
eventually their algebra, in the sequel we focus on the four-
dimensional case. It is convenient to rewrite the Lagrangian
action (\ref{fcol1}) for $D=4$ in first-order form. For subsequent
purpose, we replace the tensor fields $A_{\alpha \beta
}^{\;\;\;(\sigma )}$ by their duals with respect to the
antisymmetry indices, $\frac{1}{2}\varepsilon ^{\alpha \beta
\gamma \delta }A_{\gamma \delta (\sigma )}$, which we will still
denote by $A_{\;\;\;(\sigma )}^{\alpha \beta }$, so the Lagrangian
action in first-order form becomes
\begin{equation}
S_{0}^{\prime }\left[ A_{\;\;\;(\sigma )}^{\alpha \beta },B_{\alpha
}^{\;(\sigma )}\right] =\frac{1}{2}\int d^{4}x\left( -B_{\alpha }^{\;(\sigma
)}B_{\;\;(\sigma )}^{\alpha }+A_{\;\;\;(\sigma )}^{\alpha \beta }H_{\alpha
\beta }^{\;\;\;(\sigma )}\right) ,  \label{fcola1}
\end{equation}
by means of adding an auxiliary tensor field of degree two $B_{\alpha
}^{\;(\sigma )}$ with no symmetry, endowed with the field strength
\begin{equation}
H_{\alpha \beta }^{\;\;\;(\sigma )}=\partial _{\alpha }B_{\beta }^{\;(\sigma
)}-\partial _{\beta }B_{\alpha }^{\;(\sigma )}\equiv \partial _{\left[
\alpha \right. }B_{\left. \beta \right] }^{\;\;(\sigma )}.  \label{fcola2}
\end{equation}
It is known that auxiliary fields do not change the dynamics, and,
essentially, they do not modify the local cohomological group $H^{0}\left(
s|d\right) $ either \cite{32and12}. The gauge invariances of action (\ref
{fcola1}) are given by
\begin{equation}
\delta _{\epsilon }A_{\;\;\;(\sigma )}^{\alpha \beta }=\varepsilon ^{\alpha
\beta \gamma \delta }\partial _{\gamma }\epsilon _{\delta (\sigma )}\equiv
\stackrel{(0)}{R}_{\;\;\;(\sigma )}^{\prime \alpha \beta \;\;\;\delta (\rho
)}\epsilon _{\delta (\rho )},\;\delta _{\epsilon }B_{\alpha }^{\;(\sigma
)}=0,  \label{fcola3}
\end{equation}
with
\begin{equation}
\stackrel{(0)}{R}_{\;\;\;(\sigma )}^{\prime \alpha \beta \;\;\;\delta (\rho
)}=\delta _{\sigma }^{\rho }\varepsilon ^{\alpha \beta \gamma \delta
}\partial _{\gamma },  \label{fcola4}
\end{equation}
and $\varepsilon ^{\alpha \beta \gamma \delta }$ is the completely
antisymmetric symbol in four spacetime dimensions, $\varepsilon
^{0123}=+1$. They are off-shell first-stage reducible, where the
reducibility relations and functions read
\begin{equation}
\stackrel{(0)}{R}_{\;\;\;(\sigma )}^{\prime \alpha \beta \;\;\;\delta (\rho
)}\stackrel{(0)}{Z}_{\delta (\rho )}^{\prime \;\;\;\;(\tau )}=0,
\label{fcola5}
\end{equation}
\begin{equation}
\stackrel{(0)}{Z}_{\delta (\rho )}^{\prime \;\;\;\;(\tau )}=\delta _{\rho
}^{\tau }\partial _{\delta }.  \label{fcola6}
\end{equation}

The generators of the BRST complex and their degrees are listed
below:
\begin{equation}
\Phi ^{\Delta }=\left( A_{\;\;\;(\sigma )}^{\alpha \beta },B_{\alpha
}^{\;(\sigma )},\eta _{\alpha (\sigma )},C_{(\sigma )}\right) ,
\label{fcola7}
\end{equation}
\begin{equation}
\Phi _{\Delta }^{*}=\left( A_{\alpha \beta }^{*\;\;(\sigma
)},B_{\;\;\;(\sigma )}^{*\alpha },\eta ^{*\alpha (\sigma )},C^{*(\sigma
)}\right) ,  \label{fcola7a}
\end{equation}
\begin{equation}
\varepsilon \left( A_{\;\;\;(\sigma )}^{\alpha \beta }\right) =\varepsilon
\left( B_{\alpha }^{\;(\sigma )}\right) =\varepsilon \left( C_{(\sigma
)}\right) =0,\;\varepsilon \left( \eta _{\alpha (\sigma )}\right) =1,
\label{fcola8}
\end{equation}
\begin{equation}
\mathrm{agh}\left( \Phi ^{\Delta }\right) =0,\;\mathrm{agh}\left( A_{\alpha
\beta }^{*\;\;(\sigma )}\right) =1=\mathrm{agh}\left( B_{\;\;\;(\sigma
)}^{*\alpha }\right) ,  \label{fcola9}
\end{equation}
\begin{equation}
\mathrm{agh}\left( \eta ^{*\alpha (\sigma )}\right) =2,\;\mathrm{agh}\left(
C^{*(\sigma )}\right) =3,  \label{fcola10}
\end{equation}
\begin{equation}
\mathrm{pgh}\left( \Phi _{\Delta }^{*}\right) =0,\;\mathrm{pgh}\left(
A_{\;\;\;(\sigma )}^{\alpha \beta }\right) =0=\mathrm{pgh}\left( B_{\alpha
}^{\;(\sigma )}\right) ,  \label{fcola11}
\end{equation}
\begin{equation}
\mathrm{pgh}\left( \eta _{\alpha (\sigma )}\right) =1,\;\mathrm{pgh}\left(
C_{(\sigma )}\right) =2.  \label{fcola12}
\end{equation}
In terms of the new variables, the BRST differential $s=\delta +\gamma $ of
the free theory has the form
\begin{equation}
\delta \Phi ^{\Delta }=0,\;\delta A_{\alpha \beta }^{*\;\;(\sigma )}=-\frac{1%
}{2}H_{\alpha \beta }^{\;\;\;(\sigma )},\;\delta B_{\;\;\;(\sigma
)}^{*\alpha }=B_{\;\;(\sigma )}^{\alpha }+\partial _{\beta }A_{\;\;\;(\sigma
)}^{\beta \alpha },  \label{fcola13}
\end{equation}
\begin{equation}
\delta \eta ^{*\alpha (\sigma )}=\varepsilon ^{\alpha \beta \gamma \delta
}\partial _{\beta }A_{\gamma \delta }^{*\;\;(\sigma )},\;\delta C^{*(\sigma
)}=\partial _{\alpha }\eta ^{*\alpha (\sigma )},  \label{fcola14}
\end{equation}
\begin{equation}
\gamma \Phi _{\Delta }^{*}=0,\;\gamma A_{\;\;\;(\sigma )}^{\alpha \beta
}=\varepsilon ^{\alpha \beta \gamma \delta }\partial _{\gamma }\eta _{\delta
(\sigma )},\;\gamma \left( B_{\alpha }^{\;(\sigma )}\right) =0,
\label{fcola15}
\end{equation}
\begin{equation}
\gamma \eta _{\alpha (\sigma )}=\partial _{\alpha }C_{(\sigma )},\;\gamma
C_{(\sigma )}=0,  \label{fcola16}
\end{equation}
whereas the solution to the classical master equation for the free model
with auxiliary fields is
\begin{equation}
\stackrel{(0)}{S}=S_{0}^{\prime }\left[ A_{\;\;\;(\sigma )}^{\alpha \beta
},B_{\alpha }^{\;(\sigma )}\right] +\int d^{4}x\left( \varepsilon ^{\alpha
\beta \gamma \delta }A_{\alpha \beta }^{*\;\;(\sigma )}\partial _{\gamma
}\eta _{\delta (\sigma )}+\eta ^{*\alpha (\sigma )}\partial _{\alpha
}C_{(\sigma )}\right) ,  \label{fcola17}
\end{equation}
being understood that $s\cdot =\left( \cdot ,\stackrel{(0)}{S}\right) $.

\section{Lagrangian BRST deformation in four dimensions}

We have seen that (\ref{fcol10}) requires that the first-order
deformation of the solution to the master equation is an $s$-cocycle modulo $%
d$ at ghost number zero
\begin{equation}
\stackrel{(1)}{S}=\int d^{4}x\,a,\;s\stackrel{(1)}{S}=0\Leftrightarrow
sa=\partial _{\mu }u^{\mu }.  \label{fcola19}
\end{equation}
In four dimensions, on the one hand the local homology of $\delta
$ vanishes again for $k>3$ \cite{20and5},
\begin{equation}
H_{k}\left( \delta |d\right) =0,\;k>3,  \label{fcola19a}
\end{equation}
and, on the other hand, the cohomology of $\gamma $ is generated by $\left(
\Phi _{\Delta }^{*},\;B_{\alpha }^{\;(\sigma )}\right) $ and their spacetime
derivatives up to some finite orders, together with the undifferentiated
ghosts for ghosts $C_{(\sigma )}$. (We did not include the field strength
components (\ref{fcol2}) or their derivatives in $H\left( \gamma \right) $
because the auxiliary fields and their derivatives take over this role via
the equations of motion for $B_{\alpha }^{\;(\sigma )}$.) As discussed in
the previous section, the cocycle $a$ can be assumed to decompose in a
finite number of terms via the antighost number
\begin{equation}
a=\sum_{k=0}^{n}a_{k},\;\mathrm{gh}\left( a_{k}\right) =0,\;\mathrm{agh}%
\left( a_{k}\right) =k,  \label{fcola20}
\end{equation}
where the component of highest antighost number belongs to
$H\left( \gamma \right) $:
\begin{equation}
\gamma a_{n}=0.  \label{fcola21}
\end{equation}
Following a reasoning similar to that of Sect. 5, we infer that
$n=2m$ and
\begin{equation}
a_{2m}=\mu ^{\rho _{1}\cdots \rho _{m}}\left( \left[ \Phi _{\Delta
}^{*}\right] ,\left[ B_{\alpha }^{\;(\sigma )}\right] \right) C_{(\rho
_{1})}\cdots C_{(\rho _{m})},  \label{fcola23}
\end{equation}
with $\mathrm{agh}\left( \mu ^{\rho _{1}\cdots \rho _{m}}\right) =2m$, $%
\mathrm{pgh}\left( \mu ^{\rho _{1}\cdots \rho _{m}}\right) =0$. In the
meantime, the local form of the equation (\ref{fcola19}) projected on
antighost number $\left( 2m-1\right) $,%
\begin{equation}
\delta a_{2m}+\gamma a_{2m-1}=\partial _{\mu }v^{\mu },  \label{fcola22}
\end{equation}
induces the result that a necessary condition for the existence of
$a_{2m-1}$ is that
\begin{equation}
\mu ^{\rho _{1}\cdots \rho _{m}}\in H_{2m}\left( \delta |d\right) .
\label{fcola24}
\end{equation}
Combining (\ref{fcola19a}) with (\ref{fcola23}), it is legitimate to presume
that the expansion (\ref{fcola20}) stops after the first three terms ($%
n=2m=2 $),
\begin{equation}
a=a_{0}+a_{1}+a_{2},  \label{fcola25}
\end{equation}
where $a_{2}$ is of the form $\mu ^{\rho }C_{(\rho )}$, with $\mu
^{\rho }$ an element of $H_{2}\left( \delta |d\right) $, whose
dependence on the fields and antifields is like in
(\ref{fcola23}). After some computation, we find that $H_{2}\left(
\delta |d\right) $ is two-dimensional, where a possible choice of
its two independent and non-trivial elements is
\begin{equation}
\mu ^{\prime \rho }=-\left( \partial _{\sigma }\eta ^{*\alpha (\rho
)}\right) B_{\alpha }^{\;(\sigma )}+\eta ^{*\alpha (\sigma )}\partial
_{\sigma }B_{\alpha }^{\;(\rho )}+\varepsilon ^{\alpha \beta \gamma \delta
}A_{\alpha \beta }^{*\;\;(\sigma )}\partial _{\sigma }A_{\gamma \delta
}^{*\;\;(\rho )},  \label{fcola26}
\end{equation}
\begin{equation}
\mu ^{\prime \prime \rho }=-\left( \partial ^{\rho }\eta ^{*\alpha (\sigma
)}\right) B_{\alpha (\sigma )}+\eta ^{*\alpha (\sigma )}\partial ^{\rho
}B_{\alpha (\sigma )}+\varepsilon ^{\alpha \beta \gamma \delta }A_{\alpha
\beta }^{*\;\;(\sigma )}\partial ^{\rho }A_{\gamma \delta (\sigma )}^{*}.
\label{fcola27}
\end{equation}
Indeed, we have
\begin{equation}
\delta \mu ^{\prime \rho }=\partial _{\mu }\left( \varepsilon ^{\mu \alpha
\beta \gamma }\left( B_{\gamma }^{\;(\sigma )}\partial _{\sigma }A_{\alpha
\beta }^{*\;\;(\rho )}-A_{\alpha \beta }^{*\;\;(\sigma )}\partial _{\sigma
}B_{\gamma }^{\;(\rho )}\right) \right) ,  \label{fcola26a}
\end{equation}
\begin{equation}
\delta \mu ^{\prime \prime \rho }=\partial _{\mu }\left( \varepsilon ^{\mu
\alpha \beta \gamma }\left( B_{\gamma }^{\;(\sigma )}\partial ^{\rho
}A_{\alpha \beta (\sigma )}^{*}-A_{\alpha \beta }^{*\;\;(\sigma )}\partial
^{\rho }B_{\gamma (\sigma )}\right) \right) .  \label{fcola27a}
\end{equation}
Consequently, the last component in the first-order deformation (\ref
{fcola25}) takes the concrete form
\begin{eqnarray}
&&a_{2}=\left( -\left( c_{1}\partial _{\sigma }\eta ^{*\alpha (\rho
)}+c_{2}\partial ^{\rho }\eta _{\;\;\;(\sigma )}^{*\alpha }\right) B_{\alpha
}^{\;(\sigma )}+\right.  \nonumber \\
&&\eta ^{*\alpha (\sigma )}\left( c_{1}\partial _{\sigma }B_{\alpha
}^{\;(\rho )}+c_{2}\partial ^{\rho }B_{\alpha (\sigma )}\right) +  \nonumber
\\
&&\left. \varepsilon ^{\alpha \beta \gamma \delta }A_{\alpha \beta
}^{*\;\;(\sigma )}\left( c_{1}\partial _{\sigma }A_{\gamma \delta
}^{*\;\;(\rho )}+c_{2}\partial ^{\rho }A_{\gamma \delta (\sigma
)}^{*}\right) \right) C_{(\rho )},  \label{fcola28}
\end{eqnarray}
with $c_{1}$ and $c_{2}$ two real constants, and otherwise
arbitrary. From the
action of $\delta $ on $a_{2}$,%
\begin{eqnarray}
&&\delta a_{2}=-\gamma \left( \varepsilon ^{\alpha \beta \gamma \delta
}\left( A_{\alpha \beta }^{*\;\;(\sigma )}\left( c_{1}\partial _{\sigma
}B_{\gamma }^{\;(\rho )}+c_{2}\partial ^{\rho }B_{\gamma (\sigma )}\right)
-\right. \right.  \nonumber \\
&&\left. \left. \left( c_{1}\partial _{\sigma }A_{\alpha \beta }^{*\;\;(\rho
)}+c_{2}\partial ^{\rho }A_{\alpha \beta (\sigma )}^{*}\right) B_{\gamma
}^{(\sigma )}\right) \eta _{\delta (\rho )}\right) +  \nonumber \\
&&\partial _{\mu }\left( \varepsilon ^{\mu \alpha \beta \gamma }\left(
B_{\gamma }^{\;(\sigma )}\left( c_{1}\partial _{\sigma }A_{\alpha \beta
}^{*\;\;(\rho )}+c_{2}\partial ^{\rho }A_{\alpha \beta (\sigma )}^{*}\right)
-\right. \right.  \nonumber \\
&&\left. \left. -A_{\alpha \beta }^{*\;\;(\sigma )}\left( c_{1}\partial
_{\sigma }B_{\gamma }^{\;(\rho )}+c_{2}\partial ^{\rho }B_{\gamma (\sigma
)}\right) \right) C_{(\rho )}\right) ,  \label{fcola28a}
\end{eqnarray}
(\ref{fcola22}) for $m=1$ yields $a_{1}$:
\begin{eqnarray}
&&a_{1}=\varepsilon ^{\alpha \beta \gamma \delta }\left( A_{\alpha \beta
}^{*\;\;(\sigma )}\left( c_{1}\partial _{\sigma }B_{\gamma }^{\;(\rho
)}+c_{2}\partial ^{\rho }B_{\gamma (\sigma )}\right) -\right.  \nonumber \\
&&\left. \left( c_{1}\partial _{\sigma }A_{\alpha \beta }^{*\;\;(\rho
)}+c_{2}\partial ^{\rho }A_{\alpha \beta (\sigma )}^{*}\right) B_{\gamma
}^{(\sigma )}\right) \eta _{\delta (\rho )},  \label{fcola29}
\end{eqnarray}
up to a solution $a_{1}^{\prime }$ of the ``homogeneous'' equation
$\gamma a_{1}^{\prime }=\partial _{\mu }k^{\mu }$ and, certainly,
up to trivial irrelevant terms. The solutions of the homogeneous
equation do modify the gauge transformations, but not their
algebra, since they correspond to a vanishing $a_{2}$. Using again
the arguments from \cite{7and23}, \cite {23and6}, \cite{32and12},
\cite{36and14}, it can be shown that we can
redefine $a_{1}^{\prime }$ such that $a_{1}^{\prime }$ is a $\gamma $%
-cocycle, $\gamma a_{1}^{\prime }=0$. As $\mathrm{pgh}\left(
a_{1}^{\prime }\right) =1$, it follows that it must be linear in
the ghosts $\eta _{\delta (\rho )}$ and their spacetime
derivatives. However, on account of the relations
(\ref{fcola15}--\ref{fcola16}) we observe that neither the ghosts
nor their derivatives are non-trivial in $H\left( \gamma \right)
$, and hence we can take $a_{1}^{\prime }=0$.

At this point we are able to determine the term $a_{0}$, which is
precisely the deformed Lagrangian at order one in $g$. The
equation for $a_{0}$ follows from (\ref{fcola19}) (with $a$ in the
form (\ref {fcola25})) projected on antighost number zero:
\begin{equation}
\delta a_{1}+\gamma a_{0}=\partial _{\mu }w^{\mu }.  \label{fcola30}
\end{equation}
By means of (\ref{fcola29}), we find that
\begin{eqnarray}
&&\delta a_{1}=\gamma \left( \left( c_{1}\partial _{\sigma }B_{\alpha
}^{\;(\rho )}+c_{2}\partial ^{\rho }B_{\alpha (\sigma )}\right) B_{\beta
}^{\;(\sigma )}A_{\;\;\;(\rho )}^{\alpha \beta }\right) -  \nonumber \\
&&\partial _{\mu }\left( \varepsilon ^{\mu \alpha \beta \gamma }\left(
c_{1}\partial _{\sigma }B_{\alpha }^{\;(\rho )}+c_{2}\partial ^{\rho
}B_{\alpha (\sigma )}\right) B_{\beta }^{\;(\sigma )}\eta _{\gamma (\rho
)}\right) ,  \label{fcola31}
\end{eqnarray}
which further leads to the expression
\begin{equation}
a_{0}=-\left( c_{1}\partial _{\sigma }B_{\alpha }^{\;(\rho )}+c_{2}\partial
^{\rho }B_{\alpha (\sigma )}\right) B_{\beta }^{\;(\sigma )}A_{\;\;\;(\rho
)}^{\alpha \beta },  \label{fcola32}
\end{equation}
also up to a solution $a_{0}^{\prime }$ of the ``homogeneous'' equation $%
\gamma a_{0}^{\prime }=\partial _{\mu }l^{\mu }$ corresponding to
vanishing $a_{1}$. These solutions only modify the free action (\ref{fcola1}%
) by adding to it terms that are invariant under the gauge transformations (%
\ref{fcola3}) of the free model, $\gamma \left( \int d^{4}x\,a_{0}^{\prime
}\right) =0$. The most general form of such gauge-invariant terms is
expressed by arbitrary polynomials in the auxiliary fields $B_{\alpha
}^{\;(\sigma )}$ and their derivatives. However, these interactions are less
interesting as they merely change the field equations, but do not modify the
gauge symmetry.

Now, the first-order deformation of the solution to the master equation
\begin{equation}
\stackrel{(1)}{S}=\int d^{4}x\,a=\int d^{4}x\left( a_{0}+a_{1}+a_{2}\right) ,
\label{fcola33}
\end{equation}
where $a_{0,1,2}$ are indicated in (\ref{fcola28}),
(\ref{fcola29}) and (\ref {fcola32}), is by construction an
$s$-cocycle of ghost number zero, such that
$\stackrel{(0)}{S}+g\stackrel{(1)}{S}$ is a solution to the master
equation up to order $g$. It is essential to remark that the
interactions are not trivial, and that they truly deform the gauge
symmetry and its reducibility, since the antifields cannot be
eliminated from $a$ even by redefinitions.

Next, let us investigate the consistency of the first-order
deformation at order $g^{2}$. In this light, we invoke
(\ref{fcol11}). If we
employ the notations $\stackrel{(2)}{S}=\int d^{4}x\,b$ and $(1/2)%
\left( \stackrel{(1)}{S},\stackrel{(1)}{S}\right) =\int
d^{4}x\,\Delta $, then (\ref{fcol11}) takes the local form
\begin{equation}
\Delta =-s\,b+\partial _{\mu }f^{\mu }.  \label{fcola36}
\end{equation}
By direct computation, we get
\begin{eqnarray}
&&\Delta =\left( c_{1}\right) ^{2}\varepsilon ^{\alpha \beta \gamma \delta
}\partial _{\mu }\left( B_{\alpha }^{\;(\mu )}\left( B_{\beta }^{\;(\sigma
)}\left( \partial _{\sigma }B_{\gamma }^{\;(\rho )}\right) \eta _{\delta
(\rho )}-A_{\beta \gamma }^{*\;\;(\sigma )}\left( \partial _{\sigma
}B_{\delta }^{\;(\rho )}\right) C_{(\rho )}\right) +\right.  \nonumber \\
&&\left. \left( B_{\alpha }^{\;(\mu )}\left( \partial _{\rho }A_{\beta
\gamma }^{*\;\;(\sigma )}\right) B_{\delta }^{\;(\rho )}-A_{\alpha \beta
}^{*\;\;(\mu )}\left( \partial _{\rho }B_{\gamma }^{\;(\sigma )}\right)
B_{\delta }^{\;(\rho )}\right) C_{(\sigma )}\right) +  \nonumber \\
&&c_{2}\varepsilon ^{\alpha \beta \gamma \delta }\left( \partial _{\sigma
}B_{\gamma }^{\;(\rho )}\right) \left( c_{1}A_{\alpha \beta }^{*\;\;(\sigma
)}\left( 2\left( \partial ^{\tau }B_{\delta (\rho )}\right) C_{(\tau
)}+B_{\delta (\rho )}\partial ^{\tau }C_{(\tau )}\right) -\right.  \nonumber
\\
&&\left. c_{2}\left( 2\left( \partial ^{\tau }A_{\alpha \beta
}^{*\;\;(\sigma )}\right) C_{(\tau )}+A_{\alpha \beta }^{*\;\;(\sigma
)}\partial ^{\tau }C_{(\tau )}\right) B_{\delta (\rho )}\right) +  \nonumber
\\
&&c_{1}c_{2}\varepsilon ^{\alpha \beta \gamma \delta }\left( -B_{\delta
(\rho )}\left( \partial ^{\tau }B_{\gamma }^{\;(\rho )}\right) \partial
_{\sigma }\left( A_{\alpha \beta }^{*\;\;(\sigma )}C_{(\tau )}\right)
+\right.  \nonumber \\
&&\left( \left( \partial ^{\rho }A_{\alpha \beta }^{*\;\;(\tau )}\right)
\left( \partial _{\rho }B_{\delta (\sigma )}\right) -\left( \partial ^{\rho
}A_{\alpha \beta (\sigma )}^{*}\right) \left( \partial _{\rho }B_{\delta
}^{\;(\tau )}\right) \right) B_{\gamma }^{\;(\sigma )}C_{(\tau )}+  \nonumber
\\
&&\left( A_{\alpha \beta (\sigma )}^{*}\partial ^{\tau }B_{\gamma
}^{\;(\sigma )}-\left( \partial ^{\tau }A_{\alpha \beta (\sigma
)}^{*}\right) B_{\gamma }^{\;(\sigma )}\right) \partial ^{\rho }\left(
B_{\delta (\rho )}C_{(\tau )}\right) +  \nonumber \\
&&+\left( \partial ^{\rho }A_{\alpha \beta (\sigma )}^{*}\right) \left(
2\left( \partial ^{\tau }B_{\gamma }^{\;(\sigma )}\right) C_{(\tau
)}+B_{\gamma }^{\;(\sigma )}\partial ^{\tau }C_{(\tau )}\right) B_{\delta
(\rho )}-  \nonumber \\
&&\left( 2\left( \partial ^{\tau }A_{\alpha \beta (\sigma )}^{*}\right)
C_{(\tau )}+A_{\alpha \beta (\sigma )}^{*}\partial ^{\tau }C_{(\tau
)}\right) \left( \partial ^{\rho }B_{\gamma }^{\;(\sigma )}\right) B_{\delta
(\rho )}+  \nonumber \\
&&A_{\alpha \beta (\sigma )}^{*}\left( \partial ^{\rho }B_{\gamma
}^{\;(\sigma )}\right) \left( \partial _{\rho }B_{\delta }^{\;(\tau
)}\right) C_{(\tau )}-B_{\gamma (\rho )}\left( \partial ^{\rho }B_{\beta
(\sigma )}\right) \partial ^{\tau }\left( B_{\alpha }^{\;(\sigma )}\eta
_{\delta (\tau )}\right) +  \nonumber \\
&&\left. B_{\alpha }^{\;(\sigma )}\left( \left( \partial _{\sigma }B_{\beta
}^{\;(\rho )}+\partial ^{\rho }B_{\beta (\sigma )}\right) \left( \partial
_{\rho }B_{\gamma }^{\;(\tau )}\right) \eta _{\delta (\tau )}+\left(
\partial ^{\tau }B_{\beta (\sigma )}\right) \partial ^{\rho }\left(
B_{\gamma (\rho )}\eta _{\delta (\tau )}\right) \right) \right) +  \nonumber
\\
&&\left( c_{2}\right) ^{2}\varepsilon ^{\alpha \beta \gamma \delta }\left(
B_{\alpha }^{\;(\sigma )}\left( \partial ^{\rho }B_{\beta (\sigma )}\right)
\left( 2\left( \partial ^{\tau }B_{\gamma (\rho )}\right) \eta _{\delta
(\tau )}+B_{\gamma (\rho )}\partial ^{\tau }\eta _{\delta (\tau )}\right)
+\right.  \nonumber \\
&&\left. \left( A_{\alpha \beta (\sigma )}^{*}\partial ^{\rho }B_{\gamma
}^{\;(\sigma )}-\left( \partial ^{\rho }A_{\alpha \beta (\sigma
)}^{*}\right) B_{\gamma }^{\;(\sigma )}\right) \left( 2\left( \partial
^{\tau }B_{\delta (\rho )}\right) C_{(\tau )}+B_{\delta (\rho )}\partial
^{\tau }C_{(\tau )}\right) \right) .  \label{fcola35}
\end{eqnarray}
By inspecting the right hand-side of (\ref{fcola35}), we observe
that all the terms proportional with $\left( c_{1}\right) ^{2}$
simply reduce to a four-dimensional divergence, while those with
$\left( c_{2}\right) ^{2}$ and $c_{1}c_{2}$ cannot be written as
in (\ref{fcola36}). Then the consistency of $\stackrel{(1)}{S}$
requires that
\begin{equation}
c_{2}=0.  \label{fcola34}
\end{equation}
Under these considerations, (\ref{fcol11}) is satisfied for $%
\stackrel{(2)}{S}=0$. The remaining equations involved with the
higher-order deformations hold if we set
$\stackrel{(k)}{S}=0,\;k>2$. In addition, the constant $c_{1}$
takes any non-zero real value, and, for definiteness, will be
fixed to unity
\begin{equation}
c_{1}=1.  \label{fcola37}
\end{equation}

\section{Identification of the interacting theory}

Now, we are in the position to identify the interacting theory.
The complete deformed solution to the master equation for the
model under study, which is consistent to all orders in the
deformation parameter, reads
\begin{eqnarray}
&&\bar{S}=\int d^{4}x\left( \frac{1}{2}\left( -B_{\alpha }^{\;(\sigma
)}B_{\;\;(\sigma )}^{\alpha }+A_{\;\;\;(\rho )}^{\alpha \beta }\bar{H}%
_{\alpha \beta }^{\;\;\;(\rho )}\right) +\right.  \nonumber \\
&&\varepsilon ^{\alpha \beta \gamma \delta }\left( A_{\alpha \beta
}^{*\;\;(\sigma )}\left( D_{\gamma }\right) _{\;\;\sigma }^{\rho }-g\partial
_{\sigma }A_{\alpha \beta }^{*\;\;(\rho )}B_{\gamma }^{(\sigma )}\right)
\eta _{\delta (\rho )}+  \nonumber \\
&&\left( \eta ^{*\alpha (\sigma )}\left( D_{\alpha }\right) _{\;\;\sigma
}^{\rho }-g\left( \partial _{\sigma }\eta ^{*\alpha (\rho )}\right)
B_{\alpha }^{\;(\sigma )}\right) C_{(\rho )}-  \nonumber \\
&&\left. g\varepsilon ^{\alpha \beta \gamma \delta }A_{\alpha \beta
}^{*\;\;(\sigma )}\left( \partial _{\sigma }A_{\gamma \delta }^{*\;\;(\rho
)}\right) C_{(\rho )}\right) ,  \label{fcola42}
\end{eqnarray}
where the field strength components of $B_{\alpha }^{\;(\rho )}$ are
deformed like
\begin{equation}
\bar{H}_{\alpha \beta }^{\;\;\;(\rho )}=H_{\alpha \beta }^{\;\;\;(\rho
)}-g\left( \partial _{\sigma }B_{\left[ \alpha \right. }^{\;(\rho )}\right)
B_{\left. \beta \right] }^{\;(\sigma )},  \label{fcola43}
\end{equation}
and the ``covariant derivative'' is defined via
\begin{equation}
\left( D_{\gamma }\right) _{\;\;\sigma }^{\rho }=\delta _{\sigma }^{\rho
}\partial _{\gamma }+g\partial _{\sigma }B_{\gamma }^{\;(\rho )}.
\label{fcola44}
\end{equation}

The deformed solution (\ref{fcola42}) contains all the information
on the gauge structure of the resulting interacting theory. More
precisely, the terms of antighost number zero induce the
Lagrangian action of the coupled model,
\begin{equation}
\bar{S}_{0}\left[ A_{\;\;\;(\sigma )}^{\alpha \beta },B_{\alpha }^{\;(\sigma
)}\right] =\frac{1}{2}\int d^{4}x\left( -B_{\alpha }^{\;(\sigma
)}B_{\;\;(\sigma )}^{\alpha }+A_{\;\;\;(\rho )}^{\alpha \beta }\bar{H}%
_{\alpha \beta }^{\;\;\;(\rho )}\right) ,  \label{fcola45}
\end{equation}
while the pieces of antighost number one furnish its gauge transformations
\begin{eqnarray}
&&\bar{\delta}_{\epsilon }A_{\;\;\;(\sigma )}^{\alpha \beta }=\varepsilon
^{\alpha \beta \gamma \delta }\left( \left( D_{\gamma }\right) _{\;\;\sigma
}^{\rho }+g\delta _{\sigma }^{\rho }\left( \partial _{\tau }B_{\gamma
}^{\;(\tau )}+B_{\gamma }^{\;(\tau )}\partial _{\tau }\right) \right)
\epsilon _{\delta (\rho )}\equiv  \nonumber \\
&&\bar{R}_{\;\;\;(\sigma )}^{\alpha \beta \;\;\;\delta (\rho )}\epsilon
_{\delta (\rho )},\;\bar{\delta}_{\epsilon }B_{\alpha }^{\;(\sigma )}=0,
\label{fcola46}
\end{eqnarray}
where the new gauge generators of the tensor fields of degree three are
expressed by
\begin{equation}
\bar{R}_{\;\;\;(\sigma )}^{\alpha \beta \;\;\;\delta (\rho )}=\varepsilon
^{\alpha \beta \gamma \delta }\left( \left( D_{\gamma }\right) _{\;\;\sigma
}^{\rho }+g\delta _{\sigma }^{\rho }\left( \partial _{\tau }B_{\gamma
}^{\;(\tau )}+B_{\gamma }^{\;(\tau )}\partial _{\tau }\right) \right) .
\label{fcola46a}
\end{equation}
The elements with antighost number two that are simultaneously linear in the
ghosts for ghosts and in the antifields of the ghosts determine the
first-order reducibility functions
\begin{equation}
\bar{Z}_{\alpha (\sigma )}^{\;\;\;\;(\rho )}=\left( D_{\alpha }\right)
_{\;\;\sigma }^{\rho }+g\delta _{\sigma }^{\rho }\left( \partial _{\tau
}B_{\alpha }^{\;(\tau )}+B_{\alpha }^{\;(\tau )}\partial _{\tau }\right) .
\label{fcola47}
\end{equation}
The appearance of the terms quadratic in the antifields of the fields $%
A_{\;\;\;(\sigma )}^{\alpha \beta }$ and linear in the ghosts for ghosts
signifies that the first-order reducibility relations
\begin{equation}
\bar{R}_{\;\;\;(\sigma )}^{\alpha \beta \;\;\;\delta (\rho )}\bar{Z}_{\delta
(\rho )}^{\;\;\;\;(\tau )}\approx 0,  \label{fcola48}
\end{equation}
hold only on-shell (i.e., on the stationary surface of the field
equations deriving from the action (\ref{fcola45})), in contrast
to the free model, for
which the reducibility takes place off-shell. The absence of the antifields $%
B_{\;\;\;(\sigma )}^{*\alpha }$ emphasizes that the deformation
procedure does not endow the auxiliary fields with gauge
invariances. Meanwhile, the absence of pieces with antighost
number two that are both quadratic in the ghosts $\eta _{\delta
(\rho )}$ and linear in their antifields enables us to state that
the gauge algebra of the interacting model remains abelian. Of
course, we can always add to action (\ref{fcola45}) any polynomial
in the auxiliary fields, without further deforming the gauge
symmetry.

The added interactions do not spoil either the locality, or the
manifest Lorentz covariance, and, essentially, are non-trivial as
the terms involving antifields cannot be removed from the deformed
solution to the maser equation by adding to it trivial ($s$-exact
modulo $d$) terms.

\section{Conclusion}

To conclude with, in this paper we have investigated the
consistent Lagrangian interactions for a special class of
covariant reducible mixed-symmetry type tensor gauge fields of
degree three. In spacetime dimensions strictly greater than four
the couplings do not modify the gauge symmetry of the initial free
model, and are merely given by strictly gauge-invariant quantities
or generalized Chern-Simons terms. A privileged situation is
encountered in four spacetime dimensions, where there appear
non-trivial consistent interactions that truly deform the gauge
symmetry and the behaviour of the reducibility relations, but not
the gauge algebra. In this sense, both situations reveal the
rigidity of the original abelian gauge algebra against the
deformation procedure.

The analysis developed in this paper can be useful at the study of
introducing general interactions among covariant mixed-symmetry type tensor
gauge fields, such as those involved with integer higher spin gauge theories
(in direct or dual formulations).

\section*{Acknowledgments}

This work has been supported by MEC-CNCSIS-Romania (type-A grant
944/2002).

\end{document}